\documentclass[usedcolumn]{mn2e}
\usepackage{times}
\usepackage{graphicx,subfigure,epsf}

\def\xmm{{\it XMM-Newton\/}}
\def\etal{et al.\ }
\def\betamod{$\beta$-model}

\def \h50 {h$_{50}$}
\newcommand{\fluxunit}{{~erg~s$^{-1}$~cm$^{-2}$}\/}
\begin{document}
\title[\xmm~ observations of three high-redshift radio galaxies]{\xmm\ observations of three high redshift radio galaxies}
\author[E. Belsole \etal]{E. Belsole$^1$\thanks{E-mail:
e.belsole@bristol.ac.uk},  D. M. Worrall$^1$, M.J. Hardcastle$^1$, M. Birkinshaw$^1$ \& C.R.  Lawrence$^2$\\
$^1$ Department of Physics, University of Bristol, Tyndall Avenue,
Bristol BS8 1TL, UK\\
$^2$ California Institute of
Technology, Jet Propulsion Laboratory, 169-327, 4800 Oak Grove Drive, Pasadena,
CA, 91109, US}

\date{Accepted . Received ; in original form }
\maketitle

\begin{abstract}
We present the results of \xmm\ observations of  three high-redshift
powerful radio galaxies 3C\,184, 3C\,292 and 
3C\,322. Although  none of the sources lies in as rich  
an X-ray-emitting environment as is seen for some powerful 
radio galaxies at low redshift, the environments provide sufficient pressure 
to confine the radio lobes. The weak gas emission is 
particularly interesting for 3C\,184, where a gravitational arc 
is seen, suggesting the presence of a massive cluster. Here {\it 
Chandra\/} data complement the \xmm\ measurements by spatially
separating X-rays from the extended atmosphere, the nucleus and the small-scale radio source.
For 3C\,292 the X-ray-emitting gas has a temperature of $\sim$2 keV and luminosity of 6.5$\times10^{43}$ erg s$^{-1}$, characteristic of a poor cluster.  
In all three cases, structures where the magnetic-field strength can be estimated through 
combining measurements of radio-synchrotron and inverse-Compton-X-ray emission, are consistent with  being in a state of minimum total energy. 3C\,184 and 3C\,292 (and possibly 3C\,322) have a heavily absorbed component of nuclear emission of $N_{\rm H} \sim $ few 10$^{23}$ cm$^{-2}$.

\end{abstract}


\section{Introduction}\label{intro}
Powerful radio galaxies are visible to high redshift and hence can be used as pointers to large-scale structures. 
It has been hypothesised that powerful radio galaxies may represent a means of discovering high-redshift clusters of galaxies (e.g., Le F\`evre et al. 1996; Fabian et al. 2001) through their X-ray emission, with the possibility of placing constraints on cosmological parameters. Analysis of pointed ROSAT observations of $z>0.5$ radio galaxies and quasars in the 3CRR catalogue (Laing, Riley \& Longair 1983) gave support to this hypothesis, resulting in claims of the detection of hot ($\sim$ 5 keV), luminous ($L_{\rm X}>10^{44} h_{50}^{-2}$ ergs s$^{-1}$) gas around a number of sources (Crawford et al. 1993, Worrall et al. 1994; Crawford \& Fabian 1995, 1996; Hardcastle et al. 1998a; Hardcastle \& Worrall 1999; Crawford et al. 1999). However, because of the limited sensitivity and spectral resolution of the instruments on-board ROSAT, there was no spectral confirmation of these results, and the detected emission was spatially resolved in relatively few cases (see Hardcastle \& Worrall 2000). With the advent of sensitive X-ray telescopes, it is now possible to detect and study the environment of such galaxies as a function of cosmic time.

Several physical mechanisms are responsible for the X-ray emission from radio galaxies. Thermal emission from a hot atmosphere  surrounding a radio galaxy gives information about the large scale gas distribution and the interaction between the expanding radio source and the external medium.
Unresolved emission  from the active galactic nucleus (AGN) may consist of  thermal or non-thermal components, and probes the physical conditions near the central engine. The radio jets, lobes and hot spots can also emit detectable X-rays, by the inverse Compton (IC) and/or synchrotron processes. IC X-ray measurements can be combined with radio data to give information about particle acceleration and magnetic field strength.

Powerful radio galaxies at intermediate redshift are found in rich, cluster-like environments (Hill \& Lilly 1991) and some (but not all) low-redshift high-power objects such as Cygnus A (e.g. Smith et al. 2002; Markevitch, Sarazin \& Vikhlinin 1999) and  Hydra A (McNamara et al. 2000; Nulsen et al. 2002) also lie in rich clusters which have been studied in detail.
The X-ray/radio properties of FRII galaxies of low/intermediate
redshift have been studied with ROSAT (e.g., Carilli, Perley \& Harris 1994; Leahy \& Gizani 1999; Hardcastle \& Worrall 2000). Hardcastle \& Worrall (2000) estimated the gas pressure in the X-ray external medium and compared it to the internal pressure of the radio-emitting plasma assuming the condition of minimum total energy in the magnetic field and radiating electrons. They concluded that the majority of the sources might be underpressured with  respect to the external medium, and discussed additional contributions to the internal pressure that would avoid collapse of the radio lobes. 

\begin{table*}
\caption{Source specific details}
\label{tab:sources}
\begin{center}
\begin{tabular}{rcccccc}
\hline
 Source	& $\alpha$(h m s)& $\delta$ (o \arcmin ~\arcsec) & redshift & $N_{\rm H}^{(3)}$ & Flux density  at 178 MHz$^{(4)}$  & scale  \\
	& (J2000)	& (J2000) 			 &	    & 10$^{20}$ (cm$^{-2}$) & (Jy) & kpc/arcmin   \\
\hline
3C 184	&07 39 24.30	& +70 23 10.7	& 0.994$^{(1)}$	& 3.45 & 13.2 & 480  \\
3C 292	&13 50 41.95	& +64 29 35.4	& 0.710$^{(1)}$ & 2.17 & 10.1 & 431  \\
3C 322	&15 35 01.16 	& +55 36 51.4	& 1.681$^{(2)}$ & 1.31 & 10.1 & 508  \\
\hline
\end{tabular}
\vskip 5pt
\end{center}
\begin{minipage}{13. cm}
(1) From Nilsson \cite{Nilsson98}- (2) from Strom et al. \cite{strometal}. (3) $N_{\rm H}$ values are from Dickey \& Lockman \cite{nh} - (4) Flux density at 178 MHz is from the 3CRR catalog - Laing, Riley \& Longair \cite{LRL83}.
\end{minipage}
\end{table*}

The new generation X-ray satellites, \xmm\ and {\it Chandra\/}, are well suited for studies of high-redshift radio galaxies, the first for its unrivalled sensitivity, the second for its excellent spatial resolution. 
{\em Chandra} observations of radio galaxies and quasars at redshift $z>0.5$ have found relatively few sources in the rich gaseous environments associated with massive clusters. Most claims are based on the flux of the extended emission (e.g. Hardcastle et al. 2002; Crawford \& Fabian 2003), with only a few confirmed by spectral measurements (3C\,220.1, $kT = 5$ keV --- Worrall et al. 2001; 3C\,294, $kT = 3.5$ keV --- Fabian et al. 2003, although a large non-thermal contribution to the emission is found to be possible.). The detected atmospheres of other sources are  more typical of a group or poor cluster (e.g. Crawford \& Fabian 2003, Donahue et al. 2003). Although the very presence of edge-brightened radio lobes points to the existence of {\em some} gas in order that the lobes should be confined, a particularly rich environment is not required, and the {\it Chandra} observations seem to point in this direction.

It is interesting to note that for  {\em Chandra}-observed radio galaxies and quasars, the diffuse X-ray emission is usually  elongated in the direction of the galaxy radio-lobe axis (e.g. Donahue et al. 2003; Carilli et al. 2002, Hardcastle et al. 2002). This has been interpreted as showing that some, if not most, of the X-ray extended emission is not thermal but arises from synchrotron or IC process (see also Fabian et al. 2003). The dominant inverse-Compton-scattered radiation field is expected to be the Cosmic Microwave Background (CMB), but on small scales this may be supplemented by photons from the nucleus and radio source (Brunetti 2000).

In this paper, we present a study of high-redshift radio galaxies based on \xmm\ observations. The large collecting area of \xmm\ provides an unprecedented opportunity for detecting relatively poor clusters  around radio galaxies at $z \sim 1$, as well as emission associated with extended radio components.
Measurements of the nuclear X-ray emission help to test models which seek to unify
high-redshift radio galaxies and quasars based on orientation, assuming the presence of a central absorbing torus of gas and dust.
Thus we might expect at least moderate intrinsic absorption of the
nuclear emission from radio galaxies, although in {\it Chandra\/}
observations to date the nucleus more commonly shows low intrinsic
absorption, suggesting that emission from an inner jet is dominant
(Worrall et al. 2001). 

Here we discuss the X-ray results for three sources: 3C\,184 ($z$=0.994), 3C\,292 ($z$=0.710) and 3C\,322 ($z$=1.681), see Table \ref{tab:sources}. The sources are all radio galaxies at $ z >0.5$ selected from the 3CRR catalogue. Their selection was made on the basis of \xmm\ observing constraints,  i.e., no nearby bright X-ray sources which might enhance the background from stray light, no nearby bright stars, and high \xmm\  visibility.  They were also chosen to be sources with no ROSAT, {\it Chandra}, or \xmm\ observations made or planned at the time they were proposed, and to enhance the sensitivity at low energies we required the Galactic $N_H$  to be less than  $5 \times 10^{20}$ cm$^{-2}$. Thus, since they were not selected for intrisic source
characteristics, the X-ray results should be typical of the 3CRR radio-galaxy population at high redshift. The two at lower redshift  belong to a larger sample of 33 radio galaxies and quasars that will be observed with {\it Spitzer} in a guaranteed-time programme. 

 Throughout the paper we use a cosmology with $H_{\rm 0}$ = 70 km s$^{-1}$ Mpc$^{-1}$, $\Omega_{\rm m}$ = 0.3, $\Omega_{\Lambda}$ = 0.7. If not otherwise stated, errors are quoted at 1$\sigma$ confidence level.

\section{Observations and data preparation}
\subsection{X-ray observations}\label{observations}
3C 184 was observed with \xmm\ for a total of 113 ks between September 2001 and October 2002 (see Table \ref{tab:obssum}). The total observing time is made up of  4 separate observations, but not all \xmm\ instruments  were operative for all of the time.  In this paper we only present XMM/EPIC data  from the MOS1, MOS2 and pn cameras, which used the THIN filter, and thus we do not consider the 3C 184 observation ID 0028540701 in the following. 

3C 292 was observed in October 2002 with the MEDIUM filter for 34 ks. 

The \xmm\ observation of 3C 322 was obtained in May 2002.  The total 42980 s THIN filter observation consists of 2 exposures separated by a 30-minute interval. The first observation was stopped because of high background level. However, the second exposure displays a high background throughout, and thus data from this second exposure are not useful.

Calibrated event files for all three sources were provided by the \xmm\ Science Operations Centre (SOC). For the pn data sets we extracted single and double events (PATTERN 0 to 4), while for the MOS the canonical PATTERNs 0-12 were selected.

\begin{table*}
\caption{Summary of the XMM-Newton observations of the three sources}
\label{tab:obssum}
\begin{center}
\begin{tabular}{rccccccc}
\hline
Source	& observation ID & date 	& Filter & PN mode & Duration  &  net exposure (MOS/pn) & comments\\
	&		&		&	 &	   & (ks)	& (ks)			&	\\
\hline
3C 184	& 0028540201	& 19-09-2001	& Thin	 & NA	   & 38.9   & 32 		& MOS only; obs 1\\
	& 0028540601	& 10-03-2002	& Thin	 & EFF	   & 40.9   & 26/16.4 	&   obs 2 \\
	& 0028540701    & 18-09-2002	& NA	 & NA 	   & 15.9   &	---	& RGS only \\
	& 0028540801	& 31-10-2002	& Thin	 & EFF	   & 17.9   & 0	& high background \\
\hline
3C 292	& 0147540101	& 29-10-2002	& Medium & FF	   & 33.9   &	20/17	& ---\\
\hline
3C 322	& 0028540301	& 17-05-2002	& Thin	 & EFF	   & 42.9   &	10/6.5	& high background \\
\hline
\end{tabular}
\vskip 5pt
\begin{minipage}{15.0 cm}
The duration refers to the total duration of the observation. PN mode: FF = Full Frame; EFF= Extended Full Frame; NA= Not Applicable. Net exposure is the good time after background screening.
\end{minipage}
\end{center}
\end{table*}

\subsection{High background rejection}
\begin{figure*}
\epsfxsize 8.5cm
\epsfbox{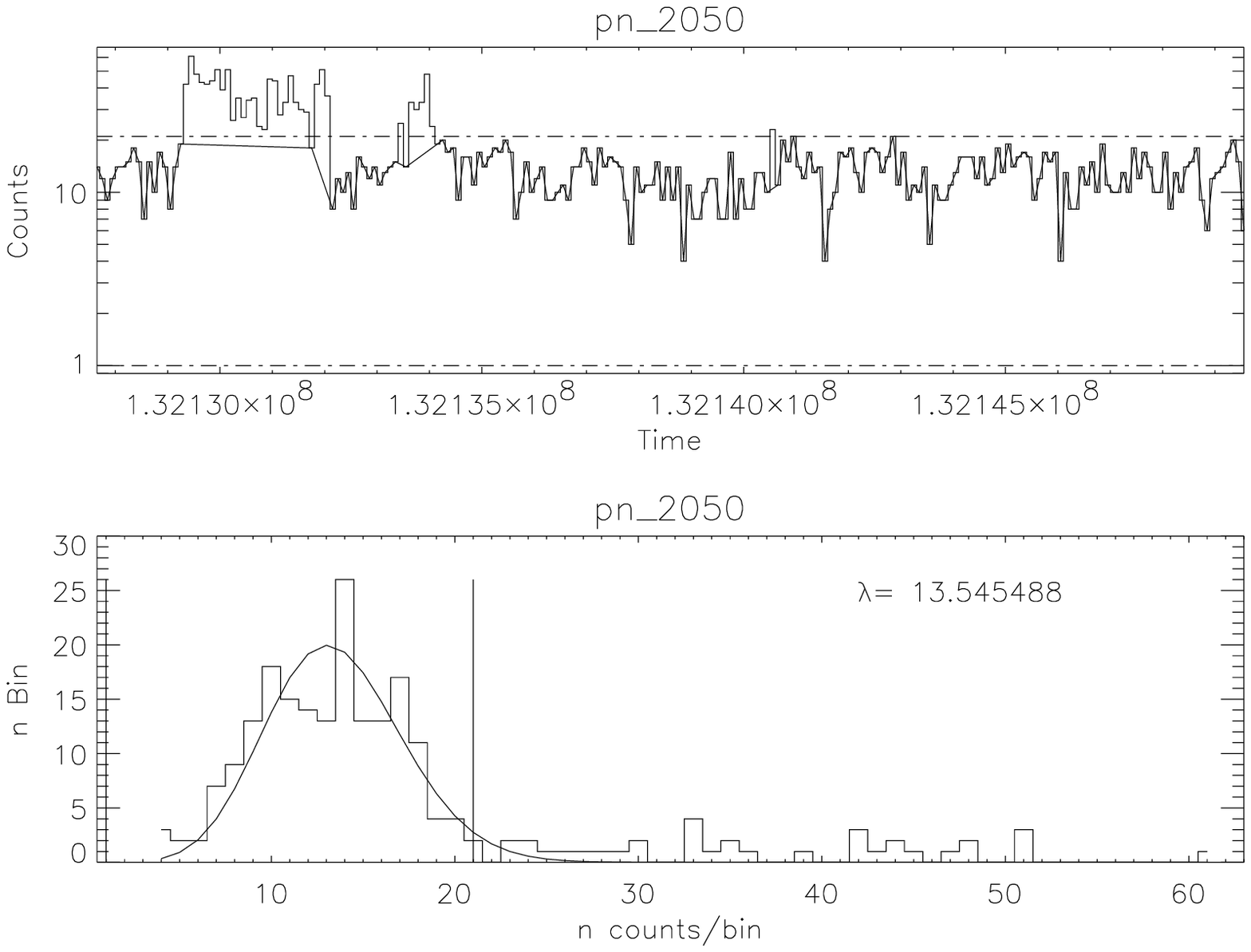}
\epsfxsize 8.5cm
\epsfbox{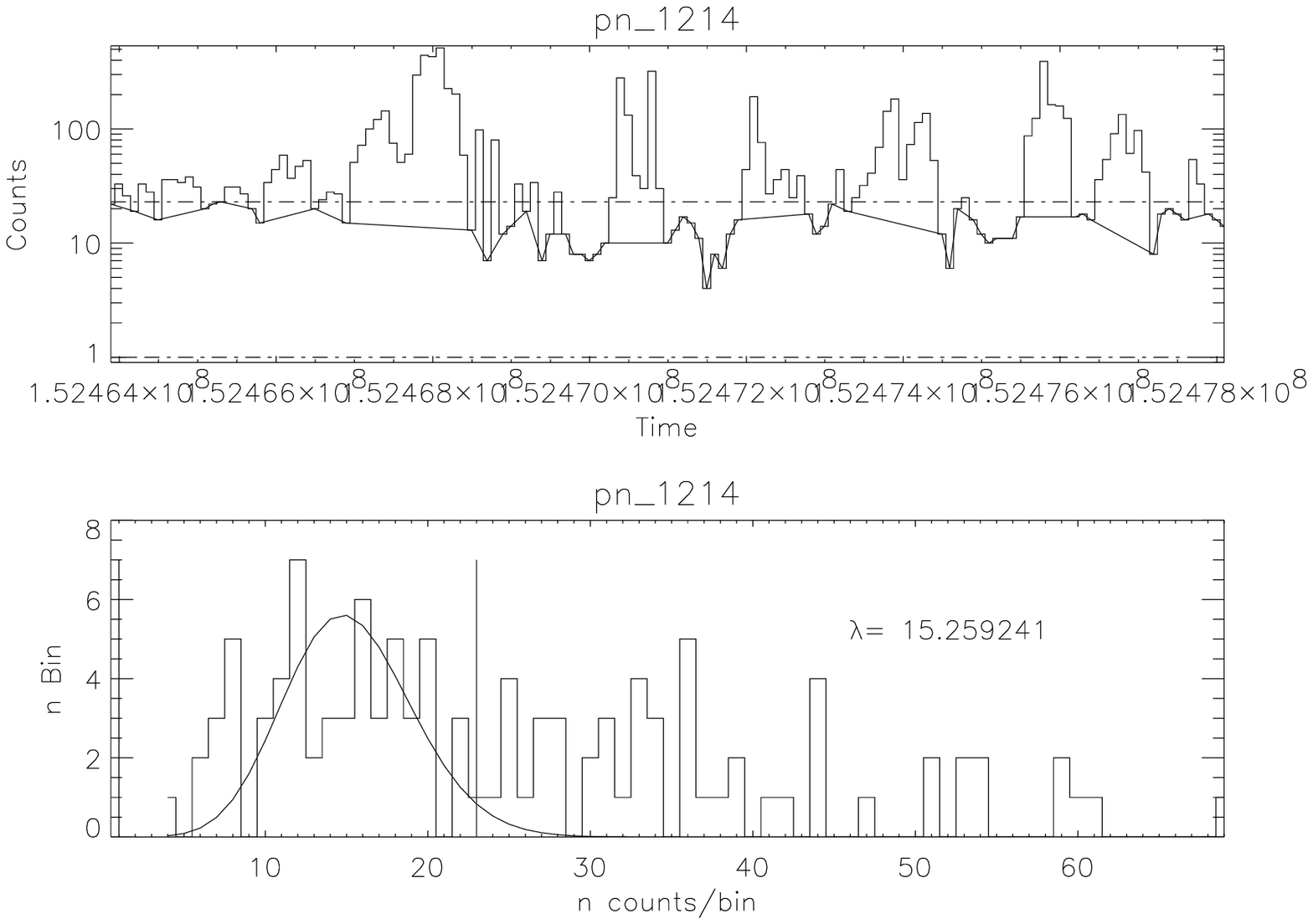}
\caption{{\em Left}: Light curve generated from the 3C\,184 event list  ID 0028540601 in the energy band 2.0-5.0 keV. The thick line shows the time intervals useful for scientific analysis. The thin line points to the time rejected by this screening. The event list was already filtered for high background by inspecting the 12-14 keV light curve. The bottom panel shows the Poisson distribution fit to the data. {\em Right}: Light curve obtained from the 12 to 14 keV photons in the pn camera for  3C 184 observation ID 0028540801. A Poisson distribution fails to describe the photon distribution (bottom panel) and the flare contamination along the whole observation is clear in the upper panel. Details in the text.}
\label{fig:lc3c184_4}
\end{figure*}
 The background component believed to be due to soft protons is highly time variable and can dominate the counts in the data file. To screen the times of highest background we applied the method described in Pratt \& Arnaud \cite{PA02}. Briefly, we extracted the MOS light curves in the 10-12 keV band and in 100 s bins, in the field of view but excluding the central CCD (only good events with FLAG=0 were used). A similar light curve was extracted for the pn but in the 12-14 keV band and by selecting double events only, because they dominate the photon-pattern distribution at these energies. We verified {\em a posteriori} that any possible variable source in the field cannot be the origin of the background fluctuation. A histogram  of each light curve was generated and fitted with a Poisson distribution, with the mean of the distribution, $\lambda$, as a free parameter. For a Poisson distribution the standard deviation of the mean is $\sigma$=$\lambda^{1/2}$ and we  normally rejected  times where the background is more than 3$\sigma$ above the mean level.  This procedure built a Good Time Intervals (GTI) file which was  applied to the event list.

In order to check the effectiveness of the high-background rejection procedure, we applied the  first GTI selection criterion as described above and  generated light curves from the cleaned event lists at lower energies. The energy bands used for this test were [0.3-2.0] keV, [2.0-5.0] keV (only single pattern events were used for the pn), [5.0-10.0] keV (MOS only), [5.0-7.0] keV (pn only, in order to avoid the instrumental lines of Cu between 7 and 10 keV in the pn camera, and double pattern only), and finally  [10.0-12.0] keV (pn, double events only).

We noticed that for all observations the cleaning with the GTI based on the higher energy band was insufficient to clean flares at lower energies.  As an example, we show in Fig. \ref{fig:lc3c184_4} (top left) the light curve obtained in the 2.0-5.0 keV band for the 3C 184 observation 0028540601, showing that screening in addition to that based on the 12-14 keV data was required.

Because we are interested in searching for extended  low-surface-brightness emission,  we require the data to have as low a background as possible while not reducing signal to noise too much by reducing the exposure time significantly. 

The final 3C 184 observation  is greatly affected by high background. The counts distribution is not  Poissonian (see Fig. \ref{fig:lc3c184_4}, right) and an inspection of the upper panel in the figure suggests that there are no time intervals which can be considered free of considerable soft proton background contamination. A more sophisticated analysis would be needed to exploit these data scientifically (modelling the high background intervals instead of rejecting them). In this paper we do not discuss this observation further.

 The net exposure times after the application of background screening are as follows:

\begin{itemize}
\item {\bf 3C\,184 observation ID 0028540201 (hereafter Obs 1)}: the GTI file was based on additional 3$\sigma$ rejection on the 5.0-10.0 light curve (MOS only). The net exposure time is 32 ks.

\item {\bf 3C\,184 observation ID 0028540601 (hereafter Obs 2)}: the GTI file was based on additional 3$\sigma$ rejection on the 2.0-5.0 keV light curve (see also Fig. \ref{fig:lc3c184_4}, left). The net exposure time is 26 ks for MOS and 16.4 ks for pn.
This observation displays some events outside the field of view (FoV)  even after a selection of {\em good} events. The ATTTSR pipeline product file implies that  this was due to an attitude problem which caused an unusually  variable pointing direction at the beginning of the observation. We rejected all affected frames and the net exposure time takes this correction into account.

\item {\bf 3C\,292}: the GTI file was based on additional 3$\sigma$ rejection on the 2.0-5.0 keV light curve. The net exposure time is 20 ks (MOS) and 17 ks (pn).

\item {\bf 3C\,322}: while the background level of the observation is high (with peaks at more than 100 counts/100 s bin) we limited the screening at the first step, thus adopting a 3$\sigma$ rejection on the 10.0-12.0 keV light curve (12.0-14.0 keV  for the pn). Even without applying a deeper screening at the low energies, the net exposure time is reduced to 10 ks and  6.5 ks for MOS and pn, respectively. The second exposure (which has the same observation number as the first one but was ``unscheduled'')  of 3C 322 displays a mean background of 600 photons in 100s bins in the high energy band. This is similar  to the level of the large background flare detected at the end of the first exposure and rejected by the method described above. This second exposure is thus rejected in total.

\end{itemize}

\subsection{Radio observations}

\begin{table*}
\caption{Properties of the radio maps used in the analysis}
\label{radiotab}
\begin{tabular}{lllrrrrl}
\hline
Source&Program&Date&Frequency&Array&Observing&FWHM&Comment\\
&ID&&(GHz)&&time (s)&(arcsec)\\
\hline
3C\,184&AM213&1987 Aug 17&4.86&A&2260&$0.52 \times 0.28$&Archive\\[2pt]
3C\,292&POOL&1987 Sep 17&1.41&B&600&$5.0 \times 5.0$&From Alexander \& Leahy (1987)\\[2pt]
&AP380&2000 Apr 27&8.46&C&2210&$2.64 \times 1.91$&Data supplied by L.M. Mullin\\[2pt]
3C\,322&AA150&1992 Dec 05&1.46&A&1330&$1.20 \times 0.97$&Archive\\
&AV133&1986 Apr 16&4.87&A&3520&$0.53 \times 0.34$&Archive\\
\hline
\end{tabular}
\vskip 5pt
\begin{minipage}{15 cm}
Column 2 gives the VLA proposal identifier for the observations. The
 FWHM of the major and minor axis of the
elliptical restoring Gaussian are given in column 7. The final column gives the origin of
the map: where this is `Archive' we obtained and calibrated data
directly from the VLA archive.
\end{minipage}
\end{table*}

The properties of the radio maps used in the analysis are summarised
in Table \ref{radiotab}. All data are from the NRAO Very Large
Array (VLA), and with
the exception of the low-resolution map of 3C\,292  the maps  were made by us
within {\sc aips} in the standard manner.

\subsection{Spatial analysis}\label{sec:radprof0}
One interest in these sources is the possibility of detecting high-redshift X-ray galaxy clusters surrounding  the radio galaxies. To search for extended emission on cluster scales, we excluded all point sources with flux greater than 2$\times$10$^{-14}$ \fluxunit ~using the pipeline product file EMSRLI. Other sources  with marginal significance in the pipeline output  were also masked. The typical radius used to excise point sources is 30 arcsec.~ For this analysis, we used event lists corrected for vignetting using the SAS task {\em evigweight}. Given the angular size and on-axis locations of the three sources, vignetting  reduces the 1.5 keV flux  by  5 per cent\footnote{Vignetting depends slightly on the camera. Here all values refer to the pn. For more details refer to the XMM User Manual - http://xmm.vilspa.esa.es/external/xmm\_user\_support/documentation/

\noindent /uhb/node19.html} 
at a distance of 2 arcmin and by 13 per cent at 4 arcmin, the largest distance from the centre considered in our analysis, in the case of 3C\,292. We minimised particle contamination by limiting the spectral range for the radial profile analysis to 0.2-2.5 keV. This selection criterion should also reduce the level of contamination of extended emission by the wings of the central point source, if the source displays a hard spectrum (see Sec. \ref{sec:3c184}). To check the consistency of our results and the adopted local background for 3C 184, we also performed a radial analysis by ignoring the vignetting correction.  The final results are in agreement.

In order to separate the pointlike and extended components of the sources, we need to include a model of the Point Spread Function (PSF). We adopted the model of the on-axis PSF described in Ghizzardi \cite{ghizzardi01}, taking into account energy dependence. All profiles are fitted (separately for each camera) with a pointlike model convolved with the PSF. A second $\beta$-model component is added when required. Such a model is a good physical description of hot gas in hydrostatic equilibrium in the potential well of a cluster (e.g Sarazin 1986).

\subsection{Spectral analysis}\label{sec:spectra0}
 Spectra were extracted from the weighted events lists, because this allows us to use the central response file and effective area for each camera.  In order to check  our results we also performed a complementary analysis on the unweighted counts, taking into account the position-dependent effective area. The two analyses gave statistically consistent results. This is not surprising given the small detector area occupied by all three sources. Consequently, only the spectral results obtained with the weighting method are described in what follows.
We adopted the SOC-provided response file version 7, as appropriate for the  date of each observation.

Since all three sources are small relative to the EPIC FoV, we took local  background estimates. Moreover, all spectral components used to fit the data are assumed to be absorbed by the Galactic column in the direction of each source, with the $N_{\rm H}$ values as given in Table \ref{tab:sources}.


\section{3C\,184}\label{sec:3c184}

\subsection{X-ray morphology}\label{sec:morpho1}
We show in Fig. \ref{fig:imaraw1} the EPIC image produced by adding events screened by the GTI from each camera and in the energy range 0.3 - 7.0 keV. The image does not take into account the vignetting correction or the exposure time, and only Obs 2 is displayed.
There are many  point sources and some possible extended sources in the FoV of the pointing. However, it is difficult, from visual inspection, to say if the nature of the 3C\,184 X-ray emission is extended or point-like. We produced an adaptively smoothed image, and the inner region of this is shown in Fig. \ref{fig:imasmooth1}. We extracted separately for each camera an image of the counts in the energy band 0.3 - 7.0 keV. We then used the SAS task {\em asmooth} to adaptively smooth the image with the fewest image counts (one of the MOS images),  and generated a smoothing template which was then applied to the other cameras. An exposure map, which takes vignetting into account, was generated  with the SAS task {\em eexpmap} and applied to each image. Finally the three camera images were added. The adaptively smoothed image of 3C\,184 is strongly dominated by point-source emission. However, some  extended emission seems to be detected out to a radius of  45 arcsec (360 kpc). The X-ray peak coincides, within 0.2 arcsec, with the core of the radio galaxy. Among the several serendipitous sources in the field, that at ($\alpha, \delta) =  (7^h39^m30.49^s;70^o23\arcmin15\farcs51$), to the north-east of 3C\,184, is noteworthy due to its proximity to 3C\,184. In a circle of radius 10 arcsec centred on this source there are 140 counts in the 0.3 - 7.0 keV band, adding data from the three EPIC cameras and the two exposures. These photon statistics are insufficient to determine the redshift of the object even if a model with lines was applicable. By comparison, 938 photons are included in a circle of radius 30 arcsec centred on the X-ray peak of 3C 184 and in the same energy band.

\subsection{Evaluation of extended X-ray emission}\label{sec:radprof1}

We performed a radial profile analysis as described in Sec. \ref{sec:radprof0}. For 3C\,184 we estimated a background in a source-centred annulus of radii 88 and 100 arcsec. There are 135, 148 and 122 net counts between 0 and 88 arcsec for the pn, MOS1, and MOS2. The two MOS1 and MOS2  observations  were merged before the extraction of the radial profile, giving MOS1 and MOS2 exposure times of 58 ks (to be compared to 16 ks of the pn camera).

We first fitted the radial profiles (separately for each camera) with a pointlike model convolved with the PSF. The radial profiles of 3C\,184 obtained with the three cameras are relatively well fitted by such a model. The fit of a point source model gives $\chi^2$/d.o.f of 6.0/8, 5.1/5 and 3.1/6 for the MOS1, MOS2, and pn, respectively. Figure \ref{fig:3c184_radprof} (Top) shows the pn camera profile fitted with a point-source model. At large distances from the centre the statistics are poor, but there is some indication of extended emission out to $\sim$2 arcmin. To quantify the degree of this extension, we added a $\beta$-model (convolved with the PSF) to the pointlike model. We selected a value of $\beta$=0.66, typical for a normal  cluster, and  fitted with a core radius, $\theta_c$, varying between 5 and 35 arcsec. 
A $\beta$ model of $\theta_c$=25 arcsec (200 kpc) plus a pointlike component is a reasonable description of the data (see Fig. \ref{fig:3c184_radprof}, bottom, where the MOS1 profile is shown), and gives a $\chi^2$/d.o.f. of 2.4/6 and 0.7/4 for MOS1 and pn respectively. The $F$-test probability of improvement in the fit occurring  by chance  is 6.4 per cent for MOS1 and  5.5 per cent for the pn, and so the \betamod\ has only  marginal significance. 
The central intensity of the \betamod\ is $I_0 = (2.3\pm1.5)\times10^{-2}$ pn-cts arcsec$^{-2}$. We estimated that within a radius of  88 arcsec, the detection radius, the extended emission contributes 69$\pm37$ (MOS1, 1$\sigma$ errors) and  67$\pm43$ (pn) counts. We computed the expected flux (and luminosity) from the pn camera for a range of temperature, by simulating the X-ray spectrum using the pn response matrix. In the energy range used for the radial profile analysis (0.2 - 2.5 keV), the unabsorbed flux depends little on temperature, and  in the temperature range between 1.5 and 5 keV the bolometric luminosity is $L_{\rm X} = 5.9\times10^{43}$ ergs s$^{-1}$ within the spatial detection radius. If we adopt a temperature of 3.5 keV (see Sec. \ref{sec:spectra1}), the cluster is underluminous by a factor of 6 with respect to the $L_{\rm X}-T$ relation of Vikhlinin et al. (2002) for clusters at $z > 0.4$.

In order to estimate the maximum  external pressure from the gaseous environment (assuming the model described above is an adequate representation), we calculate  the proton density of the gas, assuming hydrostatic equilibrium and isothermality (Cavaliere \& Fusco-Femiano 1976), for a temperature of 3.5 keV.

 We obtain a central density of 2.4$\times10^3$ m$^{-3}$ and a central pressure $3.1\times10^{-12}$ Pa. The external pressure at the radio lobes (6 arcsec) is   $\sim3.0\times10^{-12}$ Pa.

\subsection{Spectra}\label{sec:spectra1}

The compact photon distribution for the source does not allow us to spatially separate possible different components  for spectral fitting.
On the basis of the image (see Fig. \ref{fig:imasmooth1}) we extracted a global central spectrum from a circle of radius 40 arcsec~ about the peak of the source, excluding the source at the position ($\alpha; \delta) = (7^h39^m31.5^s; +70^o23\arcmin24\farcs3)$ with a circle of radius 22.5 arcsec. This optimises the signal-to-noise for spectral analysis.
The background spectrum was accumulated in four circular regions of radius 40 arcsec and at a distance between 100 and 125 arcsec from the centre of 3C\,184. At this distance from the source the background dominates over any possible extended component and we verified this further by selecting a background extraction region in an annulus of inner radius 130 arcsec and outer radius 170 arcsec. The two results are statistically consistent and our choice minimises background over-subtraction due to an erroneous correction for vignetting of the particle component.

We fitted the data between 0.2 and 10.0 keV  using several models. A summary of the spectral fits is listed in Table \ref{tab:3c184fitsp}.
A single thermal (MEKAL) model with abundances fixed to 0.3 solar, fails to give a good fit, finding a temperature of 80 keV for a $\chi^2$ = 197/50 d.o.f. A better fit is obtained with a single power law with photon index $\Gamma=0.18$, $\chi^2$ = 87/50 d.o.f. However, the power-law index is unusually  flat, and a clear improvement is obtained by adding a second component represented by an absorbed power law. 

A model composed of a power law of spectral index $\Gamma=1.78^{+0.15}_{-0.38}$ plus an absorbed power law of spectral index $\Gamma = 1.51^{+0.13}_{-0.10}$ and intrinsic absorption $N_H=5.4^{+2.2}_{-2.1} \times10^{23}$ cm$^{-2}$ gives a goodness of fit of $\chi^2$ = 42.2/47 d.o.f (model 3). Errors are quoted at 1$\sigma$ for one significant parameter. Adopting this model, the unabsorbed flux of the soft component  in the 0.2-10.0 keV energy band  is 2.0$\times10^{-14}$\fluxunit, which corresponds to a flux density at 1 keV of 1.9 nJy (but see also Sec. \ref{sec:icmodeling}). The unabsorbed flux in the 0.2-10 keV energy range of the absorbed component  is 1.0$\times10^{-13}$\fluxunit, corresponding to a luminosity of 1.4$\times10^{44}$ erg s$^{-1}$ in our cosmology. 

We also fitted with a MEKAL  model plus an absorbed power law (model 4) which gives  $\chi^2$ = 43.3/47 d.o.f. The  temperature k$T$=5.2 keV is not well constrained at its upper limit. The non-thermal absorbed component has parameters, $N_{\rm H}= 4.8^{+2.3}_{-1.8}\times10^{23}$ cm$^{-2}$ and $\Gamma=1.32^{+0.22}_{-0.29}$,  which agree with the two-power-law model (model 3). With this model the unabsorbed flux in the energy band [0.2-10.0] keV of the soft component is 1.5$\times10^{-14}$ \fluxunit and the bolometric luminosity, extrapolated up to $r_{200}$\footnote{Here the virial radius is defined as r$_{200} = h(z)^{-1} \times 3.69 \times (T/10 {\rm keV})^{1/2} h^{-1}_{50}$ Mpc, where $h(z) = (\Omega_m \times (1+z)^3 + \Lambda)^{1/2}$ and the relation is scaled from Evrard, Metzler \& Navarro~ (1996).} is  1.6$\times10^{44}$ erg s$^{-1}$, an order of magnitude too low for the best fit temperature with respect to the  $L_{\rm X}-T$ relation of Vikhlinin et al. (2002), which takes into account evolution with redshift.
However, errors are large and within the errors the fitted parameters marginally match the $L_{\rm X}-T$ relation. The unabsorbed flux of the hard absorbed component is 1.1$\times10^{-13}$ \fluxunit. We note that the residuals around 3.2 keV (see Fig. \ref{fig:3c184_centralsp} bottom) may be partially due to  neutral Fe-line emission arising from the torus. However, the addition of a Gaussian component to represent the line is not statistically required.  

Both two-component models describe the data well. The bolometric luminosity of the thermal X-ray spectral component  is about  a factor of two higher than the luminosity derived from the $\beta$-model in the previous section.  If we consider that within the 40 arcsec  region of the spectral analysis the point-source emission dominates the radial profile, it is a likely possibility that thermal emission from the cluster-like environment and non-thermal emission from a radio-related soft component both contribute to the soft X-ray spectrum.

By the time of writing, the {\em Chandra} observation of 3C\,184 had become public. In this paper we do not aim to fully describe the {\em Chandra} data. However, they help to discriminate between  models 3 and 4. We performed the analysis of the 20 ks Chandra observation in a standard way using the CIAO analysis software.
The {\em Chandra} spatial resolution allows us to separate spatially the emission from the north-western lobe and the nucleus (see figure \ref{fig:chandraima}).  We detect 5 photons spatially coincident with the NW lobe and separated from the emission of the nucleus. Some of the emission from the core looks diffuse, but no extended emission on scales larger than $\sim$5 arcsec (40 kpc) is detected. The peak of the X-ray emission in the \xmm\ and {\em Chandra} data coincides to within 0.1 arcsec.  The core of the radio galaxy, as detected at 1.4 GHz is shifted by 0.8 arcsec with respect to the peak of the X-ray emission, as shown in Figure \ref{fig:chandraima}. This is more than the aspect error expected for {\em Chandra}, which should be less than 0.6 arcsec. The radio image attitude error is of order 0.1 arcsec. Although we cannot rule out a possible physical origin for the observed shift between the X-ray and radio cores, the point-like nature of both emission makes it likely that they are related. 

 We extracted the spectrum of the core selecting a circle of radius 3.8 arcsec, which includes the emission from the nucleus and the radio lobe. There are only 52 counts in the energy range 0.4-7.0 keV, too few to place strong constraints. Nevertheless, we fitted the 0.4-7.0 keV spectrum (grouped to 5 counts per bin) using the Cash statistics with a power law plus an absorbed power law.
The best fit gives: $\Gamma_s = 1.5\pm1.0$, intrinsic absorption $N_{\rm H} = (9\pm7)\times10^{23}$ cm$^{-2}$ and hard component photon index $\Gamma_h = 3.8\pm2.3$.
The normalisation of the soft component is much lower than suggested from model 3. Assuming the core has not varied between the \xmm\ and Chandra observations, this supports the existence of  a significant thermal contribution in the \xmm\ data.

We added to model 4 a power-law model with photon index and normalisation fixed to the {\em Chandra} spectral results. We then fitted a 3-component model (model 5). We found a  best-fitting temperature of 3.6 keV. The bolometric X-ray luminosity (extrapolated out to the radial profile detection radius of 88 arcsec), of the thermal component is 7.6$^{+1.7}_{-1.6}\times10^{43}$ erg s$^{-1}$. This is in excellent agreement with the luminosity of $5.9\times10^{43}$ erg s$^{-1}$ from the $\beta$-model in our radial profile fit.
The background-subtracted spectrum and folded model (Model 5) are shown in Fig. \ref{fig:3c184_centralsp}.  In order to compare the spectral values obtained for 3C\,184 with the $L_{\rm X}-T$ relation of Vikhlinin et al. (2002), we used the radial profile to extrapolate the X-ray luminosity out to r$_{200}$, which for a temperature of 3.6 keV is 900 kpc. The bolometric X-ray luminosity is then $L_{\rm X}(r_{200}) = 8.3^{+1.9}_{-1.7}\times10^{43}$ erg s$^{-1}$, which is still a factor of 5 underluminous for the temperature.
\begin{table*}
\caption{Best fit results for the 3C\,184 spectrum}
\label{tab:3c184fitsp}
\begin{center}
\begin{tabular}{l|lc|lc|ccc|c}
\hline
	 & k$T$ & Norm$_T$      & $\Gamma_s$ & Norm$_s$ &  $N_{\rm H}$ (intrinsic) & $\Gamma_h$ & Norm$_h$ &  $\chi^2$/d.o.f. \\
	 & (keV) & (10$^{-14}$ cm$^{-5}$)	&            & (ph keV$^{-1}$ cm$^{-2}$ s$^{-1}$)  &   (10$^{23}$ cm$^{-2}$)   &	&  (ph keV$^{-1}$ cm$^{-2}$ s$^{-1}$) &  \\
	&	&	& & at 1 keV & & & at 1keV& \\
\hline
model 1  & 80       & 6.74$\times10^{-5}$& ---	     & ---	& --- & --- & ---& 197.5/50 \\
model 2  & --- & ---& 0.18 &  1.88$\times10^{-6}$ & ---& ---& ---& 87.7/50 \\
model 3  & --- & ---&  $1.78^{+0.15}_{-0.38}$ & 2.94$\times10^{-6}$ & $5.4^{+2.2}_{-2.1}$ & $1.51^{+0.13}_{-0.10}$ & 3.18$\times10^{-5}$ &  42.2/47 \\
model 4 & 5.2$^{+11.4}_{-2.3}$& 3.53$\times10^{-5}$& --- & --- &  4.8$^{+2.3}_{-1.8}$ & 1.32$^{+0.22}_{-0.29}$ & 2.32$\times10^{-5}$ & 43.3/47 \\
model 5 & 3.6$^{+14.1}_{-1.8}$& 2.27$\times10^{-5}$& 1.5(fix) & ~~1.2$\times10^{-6}$(fix) &  4.9$^{+2.2}_{-1.6}$ & 1.35$^{+0.39}_{-0.29}$ & 2.35$\times10^{-5}$ & 43.1/47 \\
\hline
\end{tabular}
\vskip 5pt
\begin{minipage}{17.5 cm}
Model 1 is a MEKAL model with abundances fixed to 0.3 solar. Model 2 is a single  power law. Model 3 is a power law plus an absorbed power law. Model 4 is a MEKAL model with abundances fixed at 0.3 solar plus an absorbed power law. Model 5 is the final adopted model (see text for details). It consists of model 4 plus an additional  power law ($\Gamma$=1.5) describing the nuclear emission detected with {\em Chandra}. All models are absorbed by the Galactic absorption  set to 3.45$\times$10$^{20}$ cm$^{-2}$ (Dickey \& Lockman, 1990). Errors are quoted at 1$\sigma$ for one interesting parameter.
\end{minipage}
\end{center}
\end{table*}



\section{3C\,292}

\subsection{X-ray morphology}\label{sec:morpho2}

\begin{table*}
\caption{Summary of the spectral fit results of 3C\,292}
\label{tab:3c292spresults} 
\begin{center}
\begin{tabular}{l|cccccccc}
\hline
region  & $\Gamma_s$  	& Norm$_s$ 				& $N_{\rm H}$ 		    &  $\Gamma_h$ & Norm$_h$ 			   & k$T$ & Norm$_{\rm T}$ & $\chi^2$/d.o.f. \\
	&		&(ph keV$^{-1}$ cm$^{-2}$ s$^{-1}$)	& (10$^{23}$ at cm$^{-2}$)  & 	    	  & (ph keV$^{-1}$ cm$^{-2}$ s$^{-1}$)&	(keV)	& (10$^{-14}$ cm$^{-5}$)	&	\\
\hline
core    & 2.6$_{-1.1}^{+1.5}$  & 2.04$\times10^{-6}$ & 2.9$_{-1.3}^{+1.8}$ & 2.20$_{-0.23}^{+0.26}$ & 5.96$\times10^{-5}$ & --- & --- & 14.5/19 \\

lobes & $1.88\pm0.26$ & 6.58$\times10^{-6}$ &  --- & --- & --- & ---& ---& 42.9/45 \\

cluster & --- & --- & 4.8$^*$ &  1.8$^*$  & 8.25$\times10^{-5}$ & 2.19$^{+3.12}_{-0.85}$ & 5.87$\times10^{-5}$ & 10.8/12 \\
\hline
\hline
\end{tabular}
\vskip 5pt
\begin{minipage}{17.0 cm}
{Subscript $s$ and $h$ are  for soft and hard component respectively. Errors are at 90 per cent confidence level except for the external medium where they are at 1$\sigma$. $^*$ The parameters were fitted, but were allowed to vary only within  90 per cent uncertainty range found for the fit of the core spectrum alone.}
\end{minipage}
\end{center}
\end{table*}

Figure \ref{fig:imaraw2} shows the \xmm\ count image of 3C\,292. For this source extended emission is evident, and displayed even more clearly by the smoothed image (Fig. \ref{fig:imasmooth2}), which was obtained as for 3C\,184. In that figure we also show contours of the radio map.
The radio emission of 3C\,292 is elongated in the same sense as the X-ray emission but the correspondence is not one-to-one  and the radio emission is more extended than the X-ray. Since  the resolution of the radio map at 1.4 GHz is roughly comparable to the \xmm\ PSF, we can make a direct morphological comparison. The core region, bright in X-ray, is self-absorbed at 1.4 GHz but detected at  8.4 GHz, and the position of the radio core and the X-ray peak agree to within 2 arcsec. The source south-east of the core ($\alpha=15^h50^m47^s.7$; $\delta=64^o28^{'}18\farcs8$; hereafter S1), at the far end of the southern radio lobe, seems not to be directly related to the radio galaxy (no radio emission is detected here), and it is more likely to be a background source. With 80 counts within a circle of radius 30 arcsec~around S1 it is not possible to determine its redshift. Unfortunately the proximity of another bright point source (S2) to the east of 3C 292, of ($\alpha,\delta$) = ($13^h50^m55^s.7$; $64^o28^{'}54\farcs7$) strongly contaminates the extended emission of 3C 292 via the wings of its PSF. The source, which is also detected in the radio (at 8.4 GHz), has more than 2000 EPIC counts between 0.3 and 7.0 keV in a source-centred circle of radius 30 arcsec, and is probably a background quasar or BL Lac object. By comparison, the net counts in a circle of the same radius centred on  3C 292 is of order 1000. 

Extended X-ray emission from 3C\,292 is detected out to 1.8 arcmin (578 kpc) in the north-west/south-east direction, while the more extended radio emission has a total length of 2.6 arcmin in the same direction. In the orthogonal direction, the smoothed image indicates  X-ray emission out to a radius of 38 arcsec, consistent with the 90\% encircled energy radius  of a point source. There is no direct correspondence between the radio hot-spots and X-ray emission.

\subsection{Radial profile analysis}\label{sec:radprof2}

For 3C 292 we extracted a radial profile for each camera out to a radius of 3.3 arcmin (1.4 Mpc). The background was estimated in an annulus between 3.4 and 4.1 arcmin, and point sources were excluded; in particular the point source S2 was excluded by masking with a circle of radius 60 arcsec. In the attempt  to exclude the emission related to the radio lobes from more extended emission, we also excluded two sectors, from the peak of the source, between position angle (PA) 126$^{\circ}$ and 208$^{\circ}$ and between PA=316$^{\circ}$ and PA=12$^{\circ}$. In the spatial region described above and after background subtraction there are  273, 130 and 84 counts in the pn, MOS1 and MOS2 camera respectively, in the energy range 0.2-2.5 keV used for the radial profile analysis.

We performed $\chi^2$ fits to the combination of a point-source and $\beta$-model over a grid of values of $\beta$ and core radius. The three cameras were analysed separately and then the $\chi^2$ values were  combined to obtain  best-fitting parameters  $\beta$ = 0.80$_{-0.25}^{+0.50}$, $\theta_{\rm c}$=19.7$_{-11.3}^{+22.5}$ arcsec for  $\chi^2$=3.9/14 d.o.f. (errors are quoted at 1$\sigma$ for two interesting parameters). The radial profile and  model for the pn camera are shown in Fig. \ref{fig:3c292_radprof}. The central intensity of the \betamod\ is I$_0$= 0.40$\pm0.05$ pn-cnts arcsec$^{-2}$.

We calculated the central density and pressure corresponding to the X-ray emitting environment around the radio galaxy, under the hypothesis that it is isothermal and in hydrostatic equilibrium.
Adopting a temperature of 2.2 keV (see Sec. \ref{sec:spectra2}) we obtain a central proton density of 8.5$_{-3.7}^{+7.3} \times 10^3$ m$^{-3}$ and central pressure of $6.8^{+5.8}_{-2.9} \times 10^{-12}$ Pa. At the radio lobes (75 arcsec from the centre) the pressure of the external medium is $2.5^{+0.9}_{-1.5} \times 10^{-13}$ Pa.

\subsection{Spectral analysis}\label{sec:spectra2}
We accumulated spectra in different regions in order to separate emission associated with the nucleus, lobes and cluster. Sources S1 and S2  to the south-east and east of 3C\,292  were excluded by masking with circles of radii 23 and  60 arcsec, respectively. 

The source spectra and corresponding backgrounds were extracted for each camera. The spectrum of the core region was extracted from a circle of radius 10 arcsec~ about the centre of the source  in order to limit any contribution from extended emission. The source is bright enough that this region contains  sufficient counts to determine the spectral signature of the nucleus. As a background we used an annular region of internal radius 30 arcsec (corresponding to 90 per cent encircled energy for a point source) and external radius 35 arcsec. The local background was subtracted in XSPEC and the spectra were grouped at 3$\sigma$ above the background.  The spectrum extracted from the pn camera shows two components (see Fig. \ref{fig:fig8a}): one at high energy (above 2.5 keV), which is also detected with both  MOS cameras, and a soft component. In order to be confident in the detection of this soft component, we investigated how changing the background region and/or the on-source extraction region changed our results. We tested two other  background regions: the annuli of radii [30 - 50 arcsec] and  [20 - 30 arcsec] and an alternative on-source extraction radius of 20 arcsec, and found good  agreement with the original result, which we adopt.

We performed a combined fit of the three EPIC camera spectra. The core spectrum is well fitted by a power law of photon index $\Gamma_s$= 2.6$_{-1.1}^{+1.5}$ plus an absorbed power law of intrinsic $N_{\rm H} = 2.9_{-1.3}^{+1.8}\times10^{23}$ cm$^{-2}$ and photon index $\Gamma_h = 2.2^{+0.26}_{-0.23}$ ($\chi^2$=14.5/19 d.o.f, errors are quoted at 90 per cent confidence for one interesting parameter). The unabsorbed flux in the 0.2 - 10.0 keV energy range of the absorbed component is 3.3$\times10^{-13}$ erg cm$^{-2}$ s$^{-1}$. Table \ref{tab:3c292spresults} lists a summary of the  spectral results. The power-law normalisation for the soft component corresponds to a low X-ray flux and is likely to be related to the radio emission of the core (see Hardcastle \& Worrall 1999).

The extraction region corresponding to the radio lobes was defined on the basis of the radio map (see Fig.~\ref{fig:fig8d}). Two ellipses  around the radio-lobe emission, masked by the circles used to excise S1 and S2, were used to accumulate the spectra. The nuclear region was masked with a circle of radius 30 arcsec. As a background we selected two source-masked ellipses on either side of the nucleus (as shown in Fig. ~\ref{fig:fig8d}) in order to subtract any cluster-associated  extended X-ray emission. The background-subtracted spectra of the two lobes  were fitted simultaneously, excluding channels below 0.3 keV. The spectrum and folded model are shown in Fig.~\ref{fig:fig8c}. A power-law model  gives the best-fitting result, with a photon index  $\Gamma=1.88\pm0.26$ (90 per cent confidence for a single parameter) for a $\chi^2$ = 44.1/46 d.o.f. No absorption in addition to the Galactic column density is required. We also fitted the lobe spectrum with a thermal model. In this case, the best fit gives $kT=5.3_{-2.0}^{+4.6}$ keV, and a $\chi^2$ = 48.7/46 d.o.f. when chemical abundances are fixed to 50 per cent solar, in order to account for the enrichment of thermal gas (e.g. Belsole et al. 2001). We note that some of the residuals at 1.5 and 2 keV observed in Fig. \ref{fig:fig8c} are partially taken into account by Fe L and Fe K line emission when using the thermal model. While a thermal origin of the emission from the vicinity of the radio lobes cannot be excluded, the non-thermal model gives a  better representation of the data.

Adopting the power-law model, the lobe X-ray emission has an unabsorbed flux of 4.1$\times10^{-14}$ \fluxunit ~in the broad  energy band [0.2-10.0] keV, corresponding to a luminosity of 8.6$\times10^{43}$ ergs s$^{-1}$ at the redshift of the radio galaxy, and a flux density at 1 keV of 4.1 nJy. 

After measuring the spectral characteristics of the nuclear and radio-lobe related emission, we attempted to separate spectrally  any cluster-related extended emission.  We extracted the global spectrum in the same 200 arcsec radius circular region used for the radial profile analysis, excluding the two sectors described in Sec. \ref{sec:radprof2}. For background we used an annular  region centred on the centre of 3C\,292 of inner radius 3.4 arcmin and outer radius 4.1 arcmin.

The spectrum and folded model is shown in figure \ref{fig:fig8b}. As expected from the radial-profile analysis, there are about 250 photons in the energy range 0.2-2.5 keV for the pn camera. The pn-spectrum clearly displays two components, one at high energy, and a soft component. The high energy component does not show up in the less sensitive MOS cameras which give a significant detection only below 3 keV. To be confident that the high-energy emission is from the core of the galaxy, we computed a radial profile in the energy range 3.0-7.0 keV and confirmed that the profile is well fitted by a point source model spread by the PSF.

The spectra from the three cameras were fitted simultaneously with a 2-component model comprising  a thermal model (MEKAL) and an intrinsically absorbed power law. Since the high-energy emission corresponds to the X-ray core of the radio galaxy, we allowed  the intrinsic absorption and power law to vary only within  the 90 per cent uncertainty range calculated from the best-fitting values obtained for the core spectrum (see Table \ref{tab:3c292spresults}) and left the normalisation  as a free parameter. We also notice that the soft component observed in the core spectrum with the pn camera is not included here. We justify this omission by considering that the MOS core spectrum does not show any soft emission, as shown in Fig. \ref{fig:fig8a}, while the MOS extended emission  is detected at low energy only.
The best fit gives  k$T = 2.19^{+3.12}_{-0.85}$ keV for the thermal component and $\chi^2$=10.56/12 d.o.f. when abundances were fixed to 0.3 solar. The bolometric X-ray luminosity is $L_X$ = (6.5$\pm2.8)\times10^{43}$ erg s$^{-1}$. 


\section{3C\,322}
\subsection{X-ray morphology}\label{sec:morpho3}
The short exposure after background screening and the small number of counts  for 3C\,322 makes it difficult to localise  clearly its X-ray emission (Fig. \ref{fig:3c322images}, Left).  The adaptively smoothed image suggests that its X-ray emission is extended but the radio emission extends further than the X-rays to the north, based on the location of   the  northern radio hotspot (see Fig. \ref{fig:3c322images}) as traced by the contours at 1.4 GHz. A similar behaviour is seen to the south, but to a lesser extent. While this is the highest-redshift source, and the X-ray statistics are not good enough for a detailed comparison between the radio and X-ray emission, the double nature of the X-ray emission suggests a correspondence with the radio lobes (the hotspots are towards the extreme spatial limit of the radio lobes). 

In Fig. \ref{fig:3c322images}, left, we observe that several sources lie within 1.8 arcmin of the centre of the FoV. In this area we also have an indication of an overdensity of photons not directly related to a specific source. However, the image in Fig. \ref{fig:3c322images} is not corrected for vignetting, which acts to reduce the flux at larger distances from the aimpoint of the telescope. 

In order to have a visual indication of any cluster-like extended  emission in the proximity of 3C\,322, we excised all sources using source-centred circles of radii between 30 and 60 arcsec. The adopted procedure is described in the {\em Chandra}-CIAO Analysis Threads\footnote{http://asc.harvard.edu/ciao/threads/diffuse\_emission/index.html\#SmoothData} and consists of replacing counts at the point source locations with the average Poisson noise estimated around each excised source. We then applied the algorithm {\em csmooth} in the CIAO data analysis system to the images from the three cameras separately, using the same template and taking into account the exposure map. The three adaptively smoothed images are then co-added.
The final image does not show  any extended diffuse emission in the field of view of the 3C\,322 radio galaxy.

Given the low photon statistics ($\sim$ 50 net counts in the 0.3-7.0 keV band from the whole EPIC), and the elliptical shape of the X-ray emission aligned with the radio source, we did not perform  a radial profile analysis for 3C\,322. However, under the hypothesis that all the photons corresponds to the emission from a cluster-like environment it is possible to set an upper limit on the expected luminosity. From the net count rate we estimated a flux of  at most 2.0$\times10^{-14}$ \fluxunit\ adopting a MEKAL model at a temperature of 4 keV (in order to match the $L_{\rm X} - T$ relation). The calculated bolometric X-ray luminosity is $\simeq5\times10^{44}$ erg s$^{-1}$.


\subsection{Spectra}\label{sec:spectra3}
The quality of the 3C 322 data is limited by the high background, which strongly reduces the exposure time useful for scientific analysis. In the region corresponding to the radio-galaxy emission, we detect 30, 24 and 67 counts for the MOS1, MOS2 and pn, respectively. After background subtraction, there are 32 counts from the pn camera. Despite the low counts we obtained a spectrum in an elliptical region containing the radio emission, and estimate the background in an annulus about the centre of the source of inner radius 50 arcsec and outer radius 80 arcsec, with point sources excluded. This gives 15 bins at 1$\sigma$ above the background for the pn camera. The MOS cameras together contribute 7 additional bins. The spectrum we obtained seems to show, as for the other two sources, at least two components, one being dominant at low energy (below 2 keV) and the other at high energy. 
With the small number of counts we cannot use Gaussian statistics and fit the spectrum by minimising $\chi^2$. We therefore used Cash statistics (Cash 1979), implemented in XSPEC 11.2, to fit the spectrum with several models.
A single power law gives a photon index of $\Gamma = 1.6_{-0.5}^{+0.3}$ where errors were estimated for the 1$\sigma$ confidence range. A single MEKAL model gives a temperature of 73 keV ($>$13 keV). Analogously to the other two sources, we also fitted the spectrum with 2 power laws (one absorbed) and a MEKAL model plus an absorbed power law. In the first case, we obtain $\Gamma_s$ = 3.1$_{-0.7}^{+0.9}$, $\Gamma_h$ = 2.3$_{-1.6}^{+3.1}$ and intrinsic absorption $N_{\rm H} = 3.7_{-2.8}^{+5.5}\times10^{23}$ cm$^{-2}$, where the subscripts $s$ and $h$ refer to the  soft and hard components of the spectrum. Interestingly, the thermal plus power law model gives more physically likely and better constrained parameters:  k$T$ = 1.4$_{-0.4}^{+1.0}$, $\Gamma_h$ = 2.1$_{-1.3}^{+3.3}$ and intrinsic absorption $N_{\rm H} = 3.3_{-2.2}^{+5.3}\times10^{23}$ cm$^{-2}$. The bolometric X-ray luminosity of the thermal component is in this case (1.5$\pm0.5)\times10^{44}$ erg s$^{-1}$, which is only slightly overluminous for its temperature if compared with the $L_{\rm X}-T$ relation of Vikhlinin et al. (2002). 

Because of the clear correlation between the X-ray emission and the radio emission, and because there is not  strong support for using a two component model, we adopt the  simple power-law model in what follows. The power-law slope is as expected from inverse-Compton scattering of a typical radio lobe population of electrons. In this case, the flux at 1 keV is 1.4 nJy.

\section{Inverse-Compton modelling}\label{sec:icmodeling}

The X-ray emission  can be compared to the predictions of inverse Compton (IC) models for the radio lobes. 

We used a computer code to predict the IC flux density as a function of the magnetic field strength, as described in Hardcastle, Birkinshaw \& Worrall \cite{mjh98} and Hardcastle et al.~\cite{mjh02}.

The synchrotron spectral model used here is described by a broken power-law electron energy spectrum:

\[
 N (\gamma) = \left\{ \begin{array}{ll}
		0  & \mbox{ $\gamma<\gamma_{min}$} \\
		N_0 \gamma^{-p}  &   \mbox{ $\gamma_{min} < \gamma <\gamma_{break}$} \\

		N_0 (\gamma_{break}^{d})\gamma^{-(p+d)}  &   \mbox{ $\gamma_{break} < \gamma <\gamma_{max}$} \\

		0  & \mbox{ $\gamma>\gamma_{max}$} \\
		\end{array}
	\right.
\]

\noindent where $N(\gamma) d\gamma$ gives the number density of electrons with Lorentz factors between $\gamma$ and $\gamma+d\gamma$. Models of shock acceleration predict $p\approx$ 2 for the low-energy power-law index and we adopt that value here.  Since we use the radio emission to normalise the electron spectrum, we adopt the equipartition magnetic field strength, $B_{eq}$, defined as:
\[
m_e c^2 \int_{\gamma_{min}}^{\gamma_{max}} \gamma N(\gamma) d \gamma = B_{eq}^2/2\mu_0
\]  
assuming that the equipartition is between the radiating electrons and the magnetic field only. $\mu_0$ is the permeability of free space and SI units are used.

For the three sources the parameters used to calculate the IC flux densities are summarised in Table \ref{tab:synres} together with the  predicted 1-keV CMB-IC and Synchrotron-Self-Compton (SCC) flux densities from this model.

\begin{table}
\caption{Inverse Compton modelling: parameters and results}
\label{tab:synres}
\begin{center}
\begin{tabular}{l|ccc}
\hline
		& 3C\,184 &	3C\,292 & 3C\,322\\
\hline
$\gamma_{min}$	& 10	  &	10	& 10	\\	
$\gamma_{break}$& 900	&	600	& 2000	\\
$\gamma_{max}$ 	&1.8$\times10^4$&8$\times10^5$&1.8$\times10^4$ \\
source length (arcsec)  & 2.6/1.6$^{*}$		& 137	& 20\\
source width (arcsec)	& 0.4	& 7	& 3.6\\
Flux	(Jy) / Freq (Ghz)	& 0.63/5.0& 11.01/0.178 & 11.0/0.178\\
\hline
B$_{eq}$ (nT)		& 18.6	& 1.7	& 4.3\\
IC flux  (nJy)		& 0.051	& 2.42	& 1.34 \\ 
SSC flux (nJy)		& 0.103 & 0.01	& 0.05\\
iR  flux (nJy)  	& 0.061 & ---	& ---\\
X-ray flux at 1 keV (nJy)&1.9 (0.2$^{**}$)	& 4.1	&1.4 \\
\hline
\end{tabular}
\vskip 5pt
\begin{minipage}{8.5 cm}
{\bf $B_{eq}$ is the equipartition magnetic field strength.$^*$ The source geometry consisted of 2 cylinders of radius 0.4 arcsec in length as in the Table.$^{**}$Obtained with {\em Chandra} data}
\end{minipage}
\end{center}
\end{table}

If we consider the scattering of photons from the CMB alone, then the predicted CMB/IC flux density for 3C\,184 is a factor $\sim$ 20 below the observed flux density in the lobes as measured by the \xmm\ data alone. We also computed the expected flux from a model whose additional source of scattered photons were infrared photons  from an hidden quasar. This model consists of a central point source with a flat spectrum giving flux density of 0.1 Jy between 3$\times10^{11}$ and 4$\times10^{12}$ Hz, which is similar to that found by  Meisenheimer et al. \cite{meis01} for objects at the same redshift as 3C\,184. The model follows the prescription of Brunetti (2000) and  is described in Hardcastle et al.  \cite{mjh02}.
At equipartition, we calculate a combined flux density at 1 keV from the three photon fields of 0.22  nJy (see Table \ref{tab:synres}), still insufficient to explain the X-ray flux if the \xmm\ measured flux is taken. In this model, the internal pressure of the radio lobes would be 9.1$\times10^{-11}$ Pa, which is a factor of 20 higher than the upper limit deduced for the X-ray external medium.
The measured X-ray flux of the soft component at this energy is 1.9 nJy if we adopt the two power-law model (model 3) described in Sec. \ref{sec:spectra1}. The magnetic field  would  need to be reduced by a factor of 3 to produce the measured flux density  at 1 keV by IC. 
However, we use the {\em Chandra} observation to disentangle the emission from the radio lobe and the core. We are unable to fit a spectrum from the lobe region alone, but with the count rate from the region corresponding to the north-western radio lobe we estimated a flux  at 1 keV of 0.2$\pm0.1$ nJy, a factor of 10 lower than what was found from the \xmm\ spectrum and in excellent agreement with the prediction from SSC/IC and iR photon scattering derived from the radio data. This supports further the thermal nature of most of the soft X-ray emission from the 3C\,184 spectrum, as observed by \xmm.

The X-ray flux of the 3C 292 lobes at 1 keV is less than a factor of two above the CMB/IC model predictions, supporting the equipartition assumption found for the lobes and hotspots of several other radio galaxies (e.g. Hardcastle et al. 2002, Brunetti et al. 1997). With an equipartition field the pressure in the lobes due to electrons and magnetic field is  7.9$\times10^{-13}$ Pa, which is in reasonable agreement with an average external pressure in the X-ray gas over the volume of the lobes ($2.5^{+0.9}_{-1.5} \times 10^{-13}$ Pa). A reduction in the lobe magnetic field by a factor of 2 would bring the internal and external pressure in balance.

In the case of 3C\,322, the prediction of the lobe IC X-ray emission from radio data is in excellent agreement with the measured X-ray flux at 1 keV, assuming a single power-law model of photon index $\Gamma$ =1.6 (see Sec. \ref{sec:spectra3})  despite our reservations about the spectral fitting. Although there is strong radio hotspot emission (see Fig. \ref{fig:3c322images}, insert) its  Synchrotron-Self-Compton (SSC) X-ray emission is negligible compared with the X-ray flux produced by the IC scattering of the CMB photons.

\section{Discussion and conclusions}

\subsection{The origin of the X-ray emission of 3C 184}
The combined spectro-imaging analysis of 3C\,184 indicates that half  the X-ray emission is produced by  an absorbed  component which is associated with the AGN nucleus. We see also non-thermal emission from a soft component which is likely to be radio related, at least in part.
 Other powerful radio galaxies (e.g. Hardcastle et al. 2002; Donahue et al. 2003) show soft X-ray emission associated with the position of the outer radio lobes. By comparing the \xmm\ observation with the 20 ks Chandra observation, the emission from the lobes is separated from the nuclear emission. Only 5 {\em Chandra} counts correspond to the region of the radio lobes, not sufficient for a spectral analysis. However, we estimated an X-ray flux density of 0.2$\pm0.1$ nJy at 1 keV, which is in excellent agreement with  the flux predicted by a model where the source of scattered photons is a combination of CMB and IR photons from the nucleus. This  allows us to conclude that the radio-galaxy lobe is very close to equipartition between the magnetic field and relativistic electrons. 

We have marginal evidence, from the spatial analysis, for a third component corresponding to a gaseous environment. The luminosity of the extended component estimated from the $\beta$-model fitting (with $\beta$ fixed at 0.66) is 5.9$\times10^{43}$ erg s$^{-1}$ within the detection radius of 88 arcsec, which rises to $L_{\rm X}(<r_{200}) = 6.5\times10^{43}$ erg s$^{-1}$ when we extrapolate out to the virial radius. For this luminosity to be in agreement with the  luminosity-temperature relation of Vikhlinin et al. (2002) the cluster should be at a temperature of $\sim 1.5$ keV.

The {\em Chandra} data fix the contribution of the core and lobe emission to the global spectrum extracted from the \xmm\ data. We found that most of the \xmm\ emission at low energy is thermal in nature and associated with emission from large spatial scales (at least 40 arcsec - 320 kpc). We interpret this result as the detection of a cluster-like environment around the radio galaxy. The best-fitting temperature is $kT = 3.6$ keV with a lower 1$\sigma$ limit of 1.7 keV. From the spectral best-fitting temperature, the measured bolometric X-ray luminosity ($L_{\rm X}(<r_{200}) = 8.3^{+1.9}_{-1.7}\times10^{43}$ erg s$^{-1}$, in reasonable agreement with the spatially-derived luminosity) places the object a factor of 5 below the $L_{\rm X} - T$ relation for cluster at $z > 0.4$ (Vikhlinin et al. 2002). Is this medium rich enough to confine the expansion of the radio galaxy? The average external pressure of the X-ray gas over the volume of the radio lobes is  3.0$\times10^{-12}$ Pa, a factor of 30 lower than the minimum internal pressure of the lobes (9.1$\times10^{-11}$ Pa). This is not surprising given the small physical size of the radio galaxy. 
Independent evidence in support of a cluster comes from the gravitational  arc detected with the HST (Deltorn et al. 1997) at 4.9 arcsec to the northeast of the radio galaxy (Fig. \ref{fig:3c184hst}). These authors detect an over-density of galaxies around the radio galaxy 3C\,184 and measure 11 galaxies at a redshift of $\approx$1. From the measured redshift they derived a velocity dispersion of $\sigma_v=634^{+206}_{-102}$ km s$^{-1}$. As a first approximation, we can estimate the expected temperature of the gas from the galaxy velocity (i.e. Sarazin (1986): $T_g \simeq 7\times10^7 {\rm K} ~(\sigma_v/1000$ km s$^{-1})^2$, which gives $T_g \simeq 2.4^{+1.9}_{-0.7}$, in good agreement with our spectral results. The comparison with the independent analysis of Deltorn et al. (1997) confirm that the temperature should be around 3 keV, despite the large uncertainties due to the limited statistics in our spectral analysis.

Deltorn et al.~ computed a mass enclosed within 40 $h_{70}^{-1}$ kpc (5 arcsec) of $\sim2.1\pm0.9\times10^{13} h_{70}^{-1}$ $M_{\odot}$ and derived a virial mass of 7.7 $\times 10^{14}$ $h_{70}^{-1}$ $M_{\odot}$, from the central velocity dispersion calculated for an isothermal sphere.

If we adopt  the proton density calculated in Sec. \ref{sec:radprof1}, we estimate a gas-mass within a cylinder of radius 5 arcsec integrated along the line of sight of  $\sim1.2\times 10^{11}$ $M_{\odot}$. We compared this value with the total mass calculated in the same cylinder (and scaled for our cosmology) of Deltorn et al. (1997), e.g. $\sim(2.1\pm0.9)10^{13}$ M$_{\odot}$, and found  a gas-mass to total-mass ratio of order 0.01. This is at least  a factor of 10 lower than that observed for clusters of galaxies at lower redshift.  Errors on this estimate are large, and the largest uncertainty is likely to come from the $\beta$-model parameters. The best estimate for the core-radius, seems to suggest a  rather shallow potential well with respect to other clusters at the same redshift (e.g. Vikhlinin et al. 2002), more similar to galaxy groups, rather than rich clusters. On the other hand, the optical analysis of Deltron et al. supports the existence of a large total mass around 3C\,184, and the temperature we measure also points in this direction. Since the spectral fit also implies an underlumious object, we can speculate that the cluster around 3C\,184 is somehow peculiar in its relatively high temperature but low luminosity (and low gas fraction) if compared to lower redshift galaxy clusters. However, we note that Croston et al. (2003) found a trend for galaxy groups harbouring a radio-loud source to be hotter than radio-quiet galaxy groups. 3C\,184 might show the effect of gas heating by a radio source at $z = 1$. The fact that the radio source is small and young and expanding might represent another piece of evidence in support to this picture.

\subsection{Component separation in 3C\,292}

At a redshift of $z=0.7$, 3C\,292 is the nearest source in our small sample. For this reason we can better separate the X-ray components spatially.

The spectral properties of the core of 3C 292 reveal an absorbed component which probably indicates  a hidden quasar. With the pn camera, we also detect soft emission, and the normalisation of this component (0.13 nJy) supports the interpretation to be  a soft radio-related emission from the core. The derived 0.2-10 keV flux (1.3$\times10^{-14}$ \fluxunit)  is  in agreement with the ROSAT-derived flux-flux correlation for radio/X-ray cores (Hardcastle \& Worrall 1999). 

We have evidence for the presence of extended  X-ray emission. Most of it is spatially correlated with the radio lobes, but we detected an additional extended component that we interpret as arising from a hot, cluster-like environment, and detect spatially out to a distance of 0.7 Mpc (100 arcsec) from the centre of the radio galaxy. Our best-fitting temperature, k$T$ = 2.2 keV, and  bolometric luminosity, $L_{\rm X}$ = 6.5$\times10^{43}$ erg s$^{-1}$, place the cluster slightly above, but consistent with, the $L_{\rm X} - T$ relation for clusters at $z>0.4$ (Vikhlinin et al. 2002) 

The spatial correlation between radio and X-ray emission, at the position of the radio lobes, strongly supports IC scattering as the physical process responsible for the X-ray emission aligned with the radio structure.  The larger spatial extent of the radio emission can be explained by lifetime effects.

However, some doubts arise on a pure non-thermal origin  because a thermal model is a good fit to the data, and there is some evidence for line Fe L and Fe K emission. Let us assume that the X-ray lobes emit thermally and their temperature is 5.3 keV, to be compared to a 2 keV temperature of the gaseous environment as obtained from the spectrum extracted on larger scales and by excluding the lobe X-ray emission. A possible physical interpretation in this case is that the gas surrounding the radio lobes  has been shocked by the expansion of the radio jet, as observed in the nearby source Cen A (Kraft et al. 2003).
If this is the case, we are observing this phenomenon at high redshift for the first time.
Although this effect is difficult to quantify,  some contamination from the core via the wings of the PSF could contribute to artificially increasing  the temperature, if a thermal model is adopted.
On the other hand, the adoption of  IC scattering as the physical model to explain the X-ray emission is also supported by requiring the radio plasma to be  only within a factor of 2 of equipartition.

Under the simple assumption that the cluster is described by an isothermal $\beta$-model, we  estimate that the pressure of the external medium at the distance of the radio lobes (2.5$\times10^{-13}$ Pa) and the pressure in the radio lobes (7.9$\times10^{-13}$ Pa) are of the same order, suggesting equilibrium of the radio-galaxy lobes with the external environment.

\subsection{3C 322}
The poor statistics do not allow us to constrain well the origin of the X-ray emission. We have a detection, with each of the three cameras, and we observe a correspondence between the X-ray emission and the radio emission from the galaxy. There is an interesting correlation between the southern hotspot and the X-ray southern peak of the emission, while the northern hotspot is  10 arcsec~ (85 kpc) from the northern X-ray peak. However, the estimated SSC X-ray emission for  hotspots is negligible with respect to the estimated  IC emission from the lobes. Despite our reservations concerning the spectral fitting, the good agreement between the X-ray flux estimated at 1 keV by adopting a single power-law model with $\Gamma$=1.6 and the expected IC flux from the lobes supports the IC interpretation. If we interpret the soft emission as thermal and due to a cluster environment, our spectral results would lead to a temperature (1.4 keV) and X-ray luminosity (1.5$\times10^{44}$ erg s$^{-1}$), consistent with the $L_{\rm X} - T $ relation when evolution is considered (Vikhlinin et al. 2002).
Core emission from this source is not obviously detected, and a second component to fit the spectrum is not required. However, if the hard spectrum is fitted, analogously to 3C\,292 and 3C\,184 with an absorbed power law, we obtain similar absorption and photon index as  for the other two sources.

\section*{Acknowledgments}
We are  grateful to J-L. Sauvageot for providing data analysis programs used in this work. We thank G.W. Pratt for providing the program for the light-curve data screening and J. H. Croston for adapting the radial profile extraction code to the specifics of this work. E.B. thanks M. Arnaud for helpful discussions. We are grateful to the anonymous referee for interesting suggestions which improved the manuscript. This paper  is based on observations obtained with XMM-Newton, an ESA science mission with instruments and contributions directly funded by ESA Member States and the USA (NASA). This research has made use of the  NASA's Astrophysics Data System.

\clearpage

\begin{figure}
\centering
\hspace{0cm}
\subfigure[] 
{
    \label{fig:imaraw1}
}
\hspace{0cm}
\subfigure[] 
{
    \label{fig:imasmooth1}
}
\caption{{\bf (a)}: XMM/EPIC image of 3C\,184. The image is in counts and not corrected for vignetting. {\bf (b)}: XMM/EPIC adaptively smoothed image of the 3C\,184. The image is obtained by adding the three camera images from Obs 2 smoothed separately with the same template and divided by the respective exposure maps - see text for details.}
\label{fig:3c184images} 
\end{figure}

\begin{figure}
\epsfxsize 8.5cm
\epsfxsize 8.5cm
\caption{Radial 0.2-2.5 keV profile obtained from the pn camera data of  3C 184 and fitted by a point-source model convolved with the PSF (Top). Radial profile obtained from the MOS1 camera. The dark grey lines corresponds to the respective contribution of the pointlike model and the $\beta$-model. The best-fitting, combined model is plotted in black on the data points (Bottom). The lower panels show the residuals in terms of $\chi$ .}
\label{fig:3c184_radprof}
\end{figure}

\begin{figure}
\centering
\hspace{0cm}
\subfigure[] 
{
    \label{fig:fig4a}
}
\hspace{0cm}
\subfigure[] 
{
    \label{fig:fig4b}
}
\caption{{\bf a}: Background subtracted spectrum and folded model 5 of the central 40 arcsec of 3C 184. The pn is in grey, the MOS in black. $\chi$ deviation from the model is in the bottom panel.{\bf b}: Data points are from the MOS 1 background subtracted spectrum. The continuous black line is model 5 as above. The dashed line is the MEKAL component; the dot-dashed line is the soft power law, fixed to the values obtained from the {\em Chandra} spectrum; the dotted line is the absorbed power law.}
\label{fig:3c184_centralsp} 
\end{figure}
\begin{figure}
\epsfxsize 8.5cm
\caption{{\em Chandra} image of 3C 184 and superimposed contours at 1.4 GHz. The size of the image pixel is 0.5 arcsec.}
\label{fig:chandraima}
\end{figure}
\begin{figure}
\centering
\hspace{0cm}
\subfigure[] 
{
    \label{fig:imaraw2}
}
\hspace{0cm}
\subfigure[] 
{
    \label{fig:imasmooth2}
}
\caption{{\bf (a)}: XMM/EPIC image of 3C\,292. The image is in counts and not corrected for vignetting. {\bf (b)}: XMM/EPIC adaptive smoothed image of the 3C\,292.Contours are traced form the VLA 1.4 GHz image.}
\label{fig:3c292images} 
\end{figure}
\begin{figure}
\epsfxsize 8.5cm
\caption{0.2-2.5 keV radial profile obtained from the pn camera data of  3C\,292 excluding the regions of the radio lobes. The dark grey lines corresponds to the respective contribution of the pointlike model and the $\beta$-model. The best-fitting, combined model is plotted in black on the data points. The bottom panel shows the residuals in term of $\chi$.}
\label{fig:3c292_radprof}
\end{figure}
\begin{figure}
\centering
\hspace{0cm}
\subfigure[] 
{
    \label{fig:fig8a}
}
\hspace{0cm}
\subfigure[] 
{
    \label{fig:fig8b}
}
\hspace{2cm}
\subfigure[] 
{
    \label{fig:fig8c}
}
\hspace{2cm}
\subfigure[] 
{
    \label{fig:fig8d}
}
\caption{{\bf a}: Data and  model for the core region of 3C\,292. The model is an  absorbed power law. {\bf b}: Data and model for the extended emission region, after masking the radio lobe region with two sectors (see text for details). The model here is a MEKAL of temperature 2.2 keV plus an absorbed power law. {\bf c}: Data and model for the radio lobe region. The model displayed here consists of a power law absorbed by Galactic absorption. {\bf d}: Region used to extract the lobe spectrum overlaid on the adaptively smoothed EPIC image in the energy band 0.3-7.0 keV. The white contours are obtained from the 1.4 GHz image. The black continuous ellipses are the source (lobes) region; the dashed ellipses are the background regions. Circles show the regions used to exclude point sources.  MOS data points are in black, pn in grey in (a - c).}
\label{fig:3c292multfig} 
\end{figure}
\begin{figure}
\begin{centering}
\caption{Left: XMM/EPIC image of 3C\,322. The image is in counts and not corrected for vignetting. Right: XMM/EPIC adaptive smoothed image of the 3C\,322.Contours are traced from the 1.4 GHz image. The inset shows a zoom of the same image, which shows more proecisely the X-ray/radio correlation.}
\label{fig:3c322images}
\end{centering}
\end{figure}
\begin{figure}
\epsfxsize 8.5cm
\caption{HST image and the superimposed contours of the XMM image for 3C\,184. The radio contours of the galaxy are also shown. The peak of the X-ray emission is centred on the optical counterpart of 3C 184.  The second peak in the X-ray has an optical counterpart and it is probably a background quasar. The shift between the radio and optical galaxies is of the same order of the attitude error of the HST. }
\label{fig:3c184hst}
\end{figure}

\end{document}